\documentclass[runningheads]{llncs}
\usepackage[T1]{fontenc}
\usepackage{graphicx}
\usepackage[hidelinks]{hyperref}
\usepackage[dvipsnames]{xcolor}
\usepackage{paralist}
\usepackage{algorithm}
\usepackage{algpseudocode}
\usepackage{tcolorbox}
\usepackage{listings}
\usepackage{booktabs}
\usepackage{array}

\newcommand{\toolname}{{\sc ARPaCCino}}

\newcommand{\nop}[1]{}

\begin{document}
\title{\toolname{}: An Agentic-RAG for Policy as Code Compliance}
\author{
Francesco Romeo\inst{1,2}\orcidID{0009-0006-3402-3675} \and \\
Luigi Arena\inst{1}\orcidID{0009-0008-9844-0229} \and \\
Francesco Blefari\inst{1,2}\orcidID{0009-0000-2625-631X} \and \\
Francesco Aurelio Pironti\inst{1}\orcidID{0009-0003-3183-2977} \and \\
Matteo Lupinacci\inst{1}\orcidID{0009-0000-2356-398X} \and \\
Angelo Furfaro\inst{1}\orcidID{0000-0003-2537-8918}
}
\authorrunning{F. Romeo et al.}
\institute{
University of Calabria, 87036, Rende (CS), Italy \\ 
\email{\{francesco.romeo, luigi.arena, francesco.blefari, francesco.pironti, matteo.lupinacci, angelo.furfaro\}@unical.it}
\and 
IMT School for Advanced Studies Lucca, 55100, Lucca (LU), Italy\\
\email{\{francesco.romeo, francesco.blefari\}@imtlucca.it}
}
\maketitle              % typeset the header of the contribution
\pagestyle{empty}
\begin{abstract}
Policy as Code (PaC) is a paradigm that encodes security and compliance policies into machine-readable formats, enabling automated enforcement in Infrastructure as Code (IaC) environments.
However, its adoption is hindered by the complexity of policy languages and the risk of misconfigurations.
In this work, we present \toolname{}, an agentic system that combines Large Language Models (LLMs), Retrieval-Augmented-Generation (RAG), and tool-based validation to automate the generation and verification of PaC rules.
Given natural language descriptions of the desired policies, \toolname{} generates formal \texttt{Rego} rules, assesses IaC compliance, and iteratively refines the IaC configurations to ensure conformance.
Thanks to its modular agentic  architecture and integration with external tools  and knowledge bases, \toolname{} supports policy validation across a wide range of technologies, including niche or emerging IaC frameworks.
Experimental evaluation involving a Terraform-based case study demonstrates \toolname{}'s effectiveness in generating syntactically and semantically correct policies, identifying non-compliant infrastructures, and applying corrective modifications, even when using smaller, open-weight LLMs.
Our results highlight the potential of agentic RAG architectures to enhance the automation, reliability, and accessibility of PaC workflows.

%Infrastructure as Code (IaC) is a pivotal paradigm for managing IT infrastructures by treating them as software defined through machine-readable files. IaC significantly enhances automation, scalability, reproducibility, and consistency across development and operational stages. However, Policy as Code (PaC) adoption faces challenges, including the steep learning curve associated with policy languages and the risk of policy drift in evolving infrastructures. Large Language Models (LLMs) offer significant opportunities to address these limitations, enabling AI-assisted policy generation, automated validation, and intelligent support for policy comprehension and exception management.
% We present \toolname, an Agentic RAG system capable of generating formal \texttt{Rego} rules from natural language policy descriptions and validating a given IaC architecture against the generated rules. Our system is adaptable to uncommon or emerging technologies, provided it is equipped with an appropriate domain-specific knowledge base.
\end{abstract}

\keywords{
    Policy as Code  \and 
    Agentic AI \and
    Retrieval Augmented Generation \and
    Large Language Models.
}

\section{Introduction}\label{sec:introduction}
Over the years, software and infrastructure management have become increasingly challenging due to the growing complexity and scale of systems.
To address these challenges, developers have embraced DevOps practices, which aim to reduce operational errors, accelerate provisioning, and support continuous updates throughout the development and operations lifecycle.
Within this context, Infrastructure as Code (IaC) has emerged as a standard practice. By expressing infrastructure specifications in machine-readable code, IaC enables automated provisioning, configuration, and management. This approach significantly improves automation, scalability, reproducibility, and consistency across the entire service lifecycle.

Despite its benefits, IaC remains prone to misconfigurations and security vulnerabilities when applied without sufficient expertise. To mitigate these risks, comprehensive testing and validation are often necessary before deployment.

Policy as Code (PaC) extends the Infrastructure as Code (IaC) paradigm to the definition of security and compliance policies, expressing them as formal, machine-readable rules that can be automatically validated and enforced during the provisioning process.
By integrating policy checks into Continuous Integration and Continuous Deployment (CI/CD) pipelines, PaC helps reduce human error and ensures infrastructure configurations meet security and compliance requirements early in the development lifecycle.

However, the adoption of PaC is often hindered by the steep learning curve of domain-specific policy languages and the difficulty of authoring correct, comprehensive rules—especially in dynamic and complex environments.

Recent advancements in Large Language Models (LLMs) and LLM-based techniques offer a promising solution to the limitations of current IaC and PaC practices, supporting their deeper integration into standard industry workflows.
LLMs can translate high-level policy descriptions, written in natural language, into formal, machine-readable rules suitable for automated validation of IaC configurations.

Additionally, AI agent-based workflows can enhance the generation and refinement of IaC and PaC artifacts by iteratively interacting with domain-specific tools and structured knowledge bases. In particular, Retrieval-Augmented Generation (RAG) techniques extend LLM capabilities with contextual, domain-specific knowledge, enabling accurate handling of niche or emerging technologies without the need for extensive retraining.

In this context, we present \toolname{}, an agentic system that combines a core reasoning LLM with RAG and specialized tools to automate the generation and validation of policies for IaC.
Given a natural language description of the desired policies, \toolname{} generates formal rules in \texttt{Rego} -- the policy language used by Open Policy Agent (OPA) -- then verifies and applies them to assess compliance of the provided IaC specification.

If the validation reveals non-compliance, \toolname{} can autonomously propose and apply iterative corrections to the IaC configuration until the specified requirements are satisfied. The system’s RAG module can be supplied with custom domain-specific knowledge bases, enabling its use across a wide range of technologies, including less common or emerging frameworks, provided that relevant documentation and examples are available.

The main contributions of this work can be summarized as follows:
\begin{itemize}
\item We propose a novel approach to Policy as Code generation and Infrastructure as Code validation, leveraging agentic systems that combine LLMs, RAG, and external tool integrations.
\item We implement this approach in the \toolname{} system, which generates formal policy rules from natural language, assesses infrastructure compliance, and iteratively refines configurations until policy conformance is achieved.
\item We demonstrate the effectiveness of \toolname{} through a realistic Terraform-based use case, showing its ability to autonomously retrieve domain knowledge, synthesize and verify policy rules, and validate or revise IaC specifications accordingly.
\end{itemize}

The remainder of this paper is organized as follows. Section~\ref{sec:background} provides background knowledge on Infrastructure as Code, Policy as Code, and AI agents. 
Section~\ref{sec:architecture} details the \toolname{} system architecture, including its core LLM engine and the available tools. 
Section~\ref{sec:use_case} presents a real use case involving Terraform, demonstrating the end-to-end workflow from a natural language policy description to a verified IaC definition. 
Section~\ref{sec:results} examines the results of the experimental evaluation conducted with different scenarios and LLMs.
Lastly, Section~\ref{sec:conclusions} discusses conclusions and outlines directions for future work.

Through \toolname{}, our aim is to advance the state of automated Policy as Code by leveraging agentic AI and RAG techniques to reduce developer burden, improve compliance, and support evolving infrastructure ecosystems.

\section{Background}\label{sec:background}

The provisioning of services in modern computing environments is a complex task that requires custom architectures tailored to specific use cases. These systems often consist of multiple interconnected components, such as microservices, databases, and networked elements, that increase the overall complexity and hinder maintainability.  
To manage this growing complexity, developers have adopted DevOps methodologies, aiming to reduce error rates, accelerate service provisioning, and enable continuous software delivery.

\subsection{Infrastructure as Code}
To support efficient Continuous Integration and Continuous Deployment~\cite{rahman2015synthesizing} in such environments, IaC
emerged as a foundational DevOps practice~\cite{humble2010continuous}. IaC enables programmatic provisioning, configuration, and management by  using machine-readable code~\cite{guerriero2019adoption}. This enhancing automation, reproducibility, and consistency across both development and operational phases.

To support IaC practice in DevOps, several languages, platforms and tools have been developed that allow the creation, customization, and orchestration of system components, including microservices, virtual machines, and networking layers. Popular IaC tools include Terraform~\cite{Bardin2025hashicorp} (declarative, cloud-agnostic), Ansible~\cite{ansible} (configuration-focused), and Pulumi~\cite{pulumi} (uses general-purpose languages). These tools enable the infrastructure to be versioned, tested, and deployed as application code.

Despite these advantages, IaC tools are still susceptible to misconfigurations and logic errors, which may lead to performance issues or security vulnerabilities. Over time, several solutions have been developed to test and validate the IaC infrastructure to ensure the correctness of the system before its deployment~\cite{rahman2020gang,hasan2020testing}.

More recently,  LLMs have been applied to IaC workflows to reduce manual effort and enhance reliability. LLMs can translate high-level natural language descriptions into valid infrastructure code, thus accelerating development and mitigating syntactic and semantic mistakes~\cite{hassan2025large,joshi2025review}. Some approaches also integrate automated validation and correction loops, enabling the detection and resolution of configuration errors prior to deployment~\cite{joshi2025review}.

\subsection{Policy as Code} 
PaC extends the IaC paradigm by codifying security, compliance, and operational policies into machine-readable formats. This enables automated policy enforcement throughout the software development lifecycle, ensuring continuous compliance and reducing human error~\cite{policy-as-code}. PaC integrates directly into CI/CD pipelines, facilitating automated validation of infrastructure configurations and application deployments against predefined rules, thereby shifting security left in the DevSecOps pipeline~\cite{rajapakse2022challenges}.

The de facto standard implementation is \emph{Open Policy Agent (OPA)}~\cite{opa}, an open-source general-purpose policy engine using the \texttt{Rego} declarative language to express fine-grained authorization, admission control (e.g., in Kubernetes), and data-filtering rules via API calls or as an integrated library.
Complementing this is \emph{HashiCorp Sentinel}~\cite{sentinel}, tailor-made for the HashiCorp enterprise stack (Terraform Enterprise, Vault, Consul, Nomad), featuring its own policy language, support for logical constructs and imports, and multiple enforcement levels (advisory, soft‑mandatory, hard‑mandatory). Sentinel enables proactive pre-deployment governance, enforcing policies as a prerequisite to resource provisioning.

Despite its advantages, PaC adoption can face challenges such as the steep learning curve for policy languages and managing policy drift in dynamic environments. However, the advent of AI and LLMs offers significant opportunities to overcome these limitations, enabling AI-assisted policy generation, automated validation, and intelligent support for policy comprehension.

\subsection{AI Agent}
An \emph{AI agent} is an autonomous software entity capable of reasoning about goals and executing actions to achieve specified objectives~\cite{Wooldridge2012}.  %(Figure~\ref{fig:AI_agent_schema}).
It is typically characterized by its ability to perceive the environment in which it operates, respond to environmental changes, and interact with external systems or other agents. 

\begin{figure}[htb]
    \centering
    \includegraphics[width=0.75\linewidth]{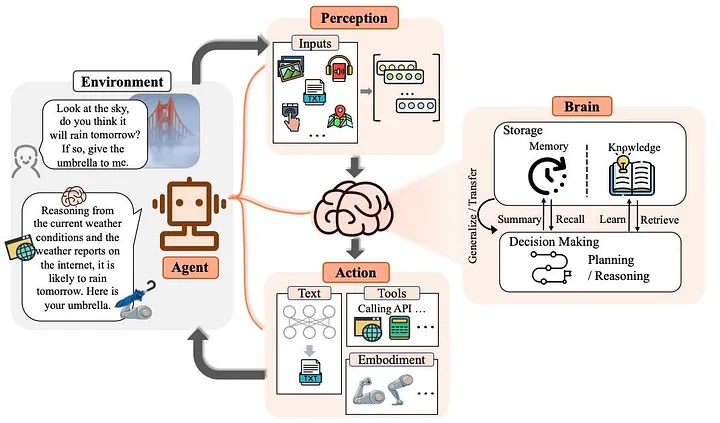}
    \caption{AI agent structure~\cite{LLM_Agents_survey}} 
    \label{fig:AI_agent_schema}
\end{figure}

An AI Agent is also provided with a \emph{memory}, useful to learn from past experiences and maintain context while addressing a task. Fig.~\ref{fig:AI_agent_schema} illustrates a schema of an AI agent structure. 

As discussed in~\cite{LLM_Agents_survey}, the emergence of LLMs marks a significant advancement in the development of intelligent agents. This evolution has led to the rise of \emph{LLM agents}, which use LLMs as the core reasoning engine to perform task decomposition, planning, and decision-making. While maintaining the reactive and interactive characteristics of traditional agents, LLM agents are augmented with the ability to invoke external tools (e.g., calculators, code interpreters, or knowledge bases) to solve domain-specific subtasks. The LLM continuously evaluates whether the task has been completed or if further tool-based refinement is required, enabling flexible and iterative problem-solving.

\nop{As discussed in~\cite{LLM_Agents_survey}, the advent of \emph{Large Language Models} (LLMs) represents a pivotal advancement in the evolution of AI agents. 
This progress has led to the rise of \emph{LLM agents}, which incorporate LLMs as their core for reasoning and decision-making. These agents retain the foundational traits of traditional AI agents while leveraging the advanced capabilities of LLMs to guide the execution flow of applications. In addition, LLM agents are augmented with the ability to interact with external tools, such as calculators or code interpreters, to tackle specialized tasks. The LLM then evaluates whether the response it produced is sufficient or if additional processing is required.
}
\subsubsection{Agentic RAG architectures}
While off-the-shelf LLMs and agent frameworks built on them offer broad utility, they often struggle with domain-specific or expert-level tasks due to the absence of embedded, up-to-date knowledge.
Although this limitation can be addressed through retraining or fine-tuning, such approaches are typically resource-intensive and time-consuming.

A more scalable and cost-effective solution is offered by the RAG paradigm~\cite{RAG}. In RAG systems, the LLM is coupled with two key components: 
\begin{inparaenum}[\itshape(i)\upshape]
    \item a \emph{repository} of domain-specific knowledge (for example, a curated collection of documents) and
    \item a \emph{retriever}  that locates relevant content from this repository to enrich the model’s input context.

\end{inparaenum}

This architecture allows LLMs to answer specialized or evolving queries by incorporating external knowledge at inference time.

A typical RAG pipeline operates as follows: an external knowledge corpus is preprocessed into manageable chunks, transformed into vector embeddings, and indexed for fast retrieval. When a user submits a query, the retriever encodes it into a vector, searches the index for the most semantically relevant chunks, and returns them. These retrieved excerpts are then combined with the user query to form an augmented prompt, which is passed to the LLM. The result is a context-aware, knowledge-informed response that extends beyond the model's original training data.

\begin{figure}[ht]
  \centering
  \includegraphics[width=0.80\linewidth]{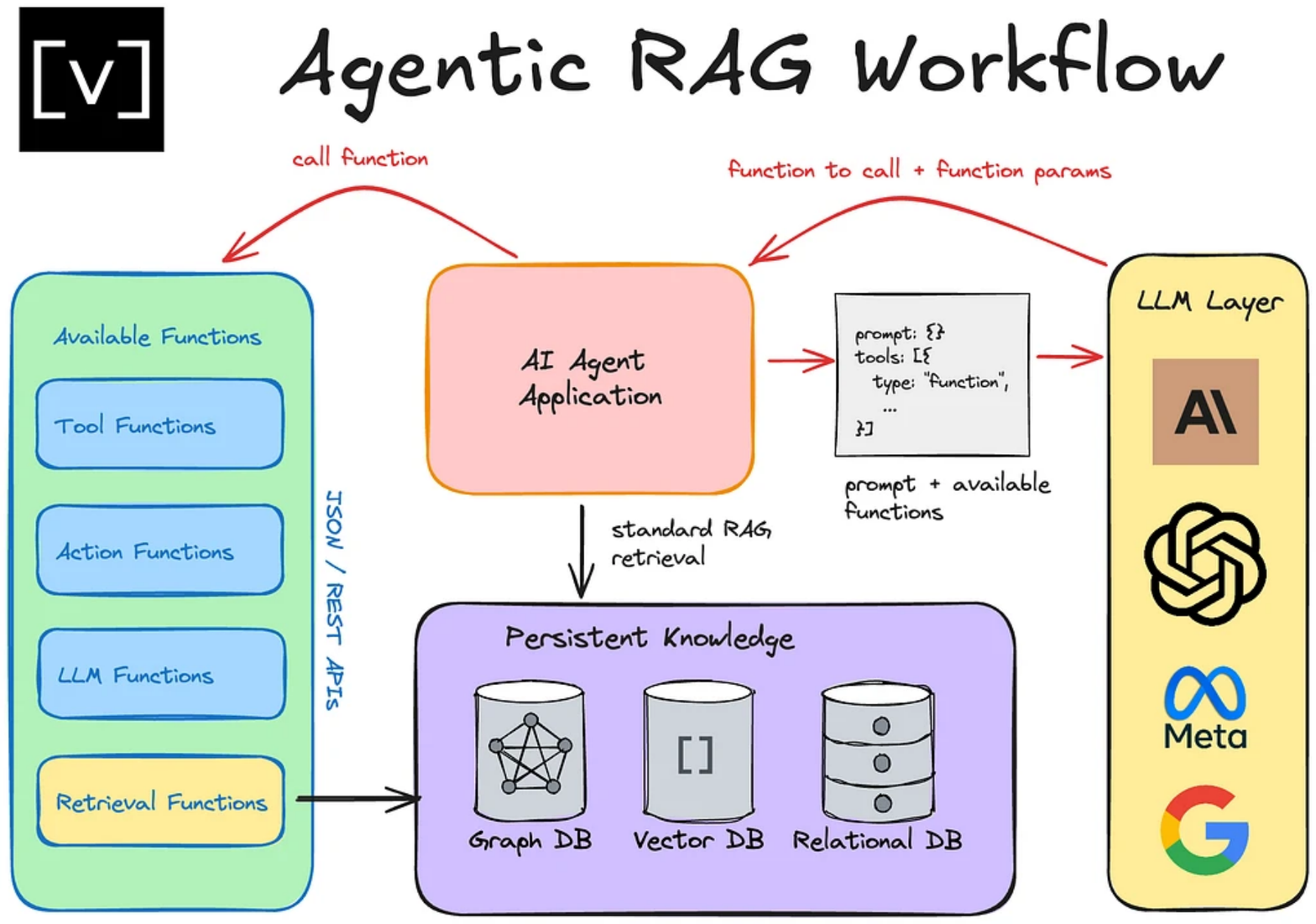}
  \caption{Agentic RAG architecture. {\footnotesize \textcopyright}  \href{https://vectorize.io/how-i-finally-got-agentic-rag-to-work-right/} {Vectorize.io} }
  \label{fig:agentic_rag_diagram}
\end{figure}

When this retrieval loop is embedded within an agent framework, enabling iterative reasoning, tool use, and multi-step workflows, the resulting architecture is referred to as \emph{Agentic RAG}. 
Fig.~\ref{fig:agentic_rag_diagram} 
provides an overview of a typical architecture of an \emph{Agentic RAG} system, illustrating the main components and their interactions during the query processing workflow.
%shows the overall architecture of such a system highlighting the interaction among the system components.
%the user query, retriever, knowledge base, LLM, and the agent’s reasoning loop.
Modern frameworks such as \textit{LangChain}~\cite{LangChain}, \textit{LlamaIndex}~\cite{LlamaIndex}, and \textit{Langroid}~\cite{Langdroid} support the rapid development of Agentic RAG systems by abstracting core components like document ingestion, embedding management, retrieval orchestration, and LLM-based decision-making.

\subsubsection{AI Agents for IaC and PaC}
Recent research has shown the growing applicability of AI techniques, in particular LLMs, in the domains of IaC and PaC.

In the context of IaC, LLMs have been successfully applied to automatically generate infrastructure definitions~\cite{palavalli2024Using,low2024Repairing,srivatsa2024Survey}.
However, these models are susceptible to well-known limitations, including \emph{hallucinations}, which may result in code that is syntactically incorrect or semantically invalid. As a result, naive applications of LLMs may introduce critical misconfigurations or deployment issues due to erroneous code in terms of both syntax and semantics.

To face these limitations, more advanced approaches have adopted LLM agents that combine reasoning with external tool integration and RAG. Some examples include the agentic architectures proposed in~\cite{lee2024LLMdriven,zhang2025DeployabilityCentric,lupinacci2025arcer}, which demonstrate the value of iterative, tool-assisted development cycles. These systems partially address the shortcomings of standard LLMs by incorporating validation, self-correction, and reasoning loops.

Similar techniques can be applied to the domain of PaC, where formal policy rules, typically expressed in languages such as \texttt{Rego}, can be generated from natural language descriptions. While some preliminary exploration in this direction exists~\cite{martinelli2024Security}, to the best of our knowledge, the system presented in this work is the first to autonomously:
\begin{inparaenum}[(i)]
    \item translate natural language policy descriptions into formal PaC rules,
    \item assess the compliance of a given IaC configuration with the generated policies, and
    \item iteratively modify the infrastructure definition to ensure full compliance with the specified requirements.
\end{inparaenum}

\section{\toolname{} Architecture}\label{sec:architecture}
\begin{figure}[ht]
    \centering
    \includegraphics[width=.9\linewidth]{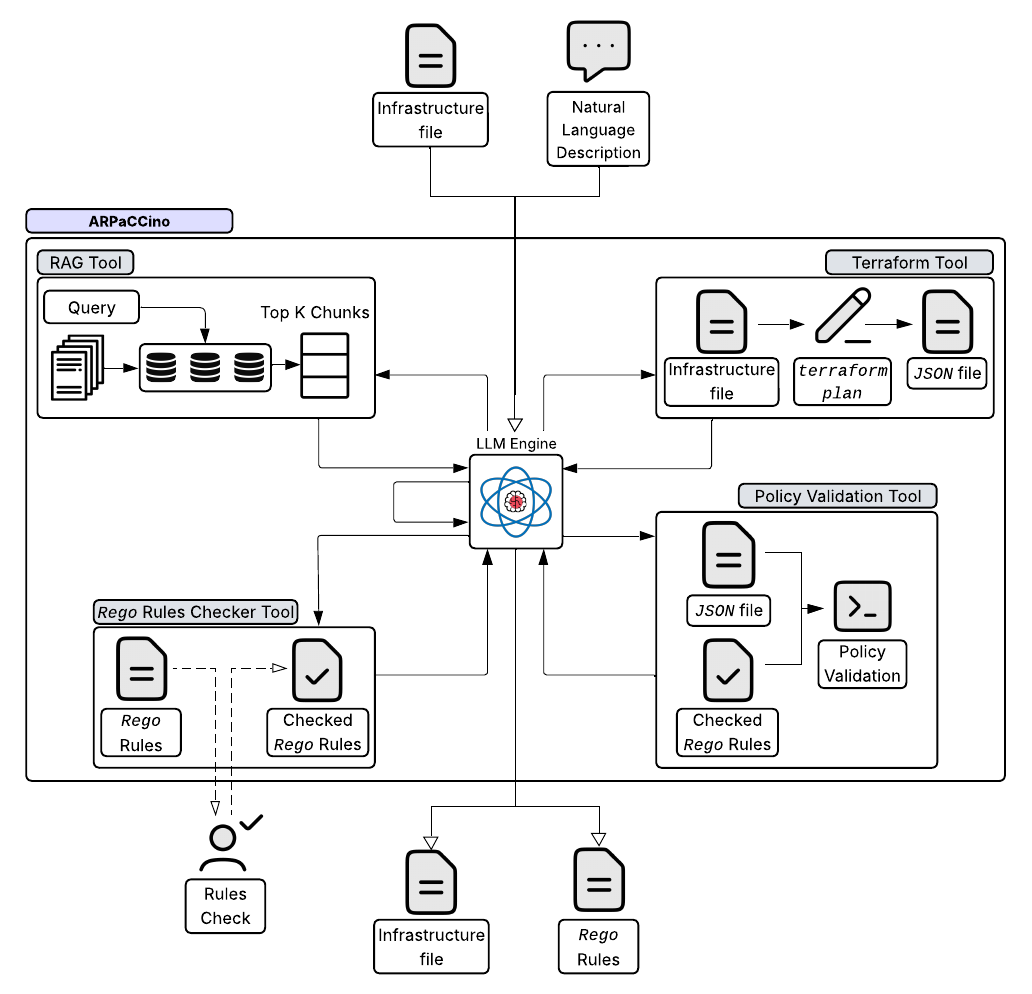}
    \caption{\toolname{}  Architecture}
    \label{fig:system_architecture}
\end{figure}

\toolname{} is an Agentic RAG system designed to translate natural language policy descriptions into formal \texttt{Rego} rules and validate IaC architectures against those policies. Leveraging the flexibility of the Agentic RAG approach, \toolname{} is able to autonomously refine the  IaC configuration until it satisfies all specified policy constraints.
%
%capable of generating formal \texttt{Rego} rules from a policy description provided in natural language and validating a given IaC architecture against the generated rules. 
%Thanks to the flexibility of the Agentic RAG approach, our system can autonomously improve the given IaC architecture until it meets the desired policy requirements.
At its core, \toolname{} consists of a reasoning engine based on an LLM, which orchestrates execution by interpreting requests, generating action plans, and invoking a suite of specialized tools. A high-level overview of the system architecture is shown in Figure~\ref{fig:system_architecture}.

%\toolname{} consists of a core LLM engine that handles the requests, elaborates an action plan, and executes it, and multiple specialized and self-contained tools that can be invoked by the engine. A high-level view of the system architecture is provided in Figure~\ref{fig:system_architecture}.

\subsubsection{RAG Tool.}
The RAG Tool provides access to domain-specific knowledge, including official documentation, the \texttt{Rego} language definition, and examples for both the Open Policy Agent (OPA) and the supported IaC frameworks.  
The LLM invokes this tool whenever domain-specific knowledge is required. The retrieved content is used to enhance the prompt, effectively extending the LLM's capabilities.  
This modularity enables \toolname{} to support uncommon or emerging technologies, provided that a structured knowledge base is supplied.

%The RAG Tool is provided with all the required knowledge about Open Policy Agent (OPA), the \texttt{Rego} language definition, and the involved Infrastructure as Code (IaC) frameworks, in the form of official documentation and useful examples.
%The LLM engine invokes this tool whenever domain-specific knowledge is required. The retrieved chunks are then used to enhance the original context.
%This eliminates the need for the LLM engine to possess prior knowledge of the frameworks involved.
%As a result, our approach can be applied even to uncommon or emerging technologies, as long as the system is supplied with an adequate knowledge base.

\subsubsection{Infrastructure Tools.}
To ensure compatibility with multiple IaC frameworks, \toolname{} leverages a set of specialized \textit{infrastructure tools}, tailored to each framework.
Those tools perform the required pre-processing on the given infrastructure definition before proceeding with the policy validation.

\subsubsection{Rule Checker Tool.}
The LLM engine, after fetching the appropriate knowledge from the RAG tool, generates the \texttt{Rego} rules corresponding to the input policy description. 
The generated rules should then be verified prior to the automatic validation and architecture improvement. %OPA provides a syntax checking functionality through the \texttt{opa check} command. However, such a command does not perform a semantic validation and is not capable of capturing flaws in the logic of the rules.
While OPA's native \texttt{opa check} command ensures syntactic correctness, it does not evaluate the semantic validity or logic of the rules.
The Rule Checker Tool addresses this limitation by incorporating feedback from an external domain expert (or oracle), who reviews and either accepts or rejects the generated rules. This step ensures the soundness of the policy prior to enforcement.
%The \texttt{Rego} Rules Checker Tool provides a semantic verification, leveraging the interaction with an external domain expert user, who accepts or discards the generated rules.

\subsubsection{Policy Validation Tool.}
The Policy Validation Tool takes as input the preprocessed infrastructure and the semantically verified \texttt{Rego} rules. 
It performs a deterministic evaluation to determine whether the infrastructure complies with the generated policies.
Based on this result, the system decides whether the current IaC specification is ready for deployment or requires further adjustments.

%the already preprocessed infrastructure definition and the generated and checked \texttt{Rego} rules, to perform an automatic (and deterministic) verification that the given infrastructure is compliant with the generated policies. 
%The results of this verification are returned to the LLM, which determines if the infrastructure is ready to be deployed or requires some adjustments.

\section{Case study}\label{sec:use_case}
To evaluate the effectiveness of \toolname{}, we present a real case study based on the widely adopted IaC framework Terraform~\cite{Bardin2025hashicorp}.
In this scenario, the knowledge base available to the RAG tool includes documentation for OPA and for the Terraform provider for ProxMox~\cite{terraformproviderproxmox}.
The infrastructure is defined using standard Terraform configuration files (\texttt{.tf}), and the system invokes the \texttt{terraform plan} command to preprocess the infrastructure. This command generates a JSON-formatted execution plan, which serves as the input to the policy validation phase.

%Using Terraform, the input infrastructure files are \texttt{.tf} textual files, while the \emph{Terraform Tool} consists of an invocation to the \texttt{terraform plan} command, which creates an execution plan and saves it in a JSON file.

% Esempio semplice: snippet terraform, descrizione policy, file rego generator, eventual terraform modification + estratto ragionamento ? 

\subsection{Expected Workflow}
% Algorithm \ref{alg:example} shows the expected workflow of \toolname{}.
% It is worth noting that the depicted scenario refers to a simplified case, as it assumes that a single tool call is sufficient to accomplish each sub-task. However, due to the agentic nature of our system, a run could require multiple calls to the same tools to achieve a satisfying result.
% Despite that, such scenarios would not alter the user experience, as the calls are autonomously handled by the agentic system.

Algorithm~\ref{alg:example} outlines the expected workflow of \toolname{}. For illustrative purposes, we assume a simplified scenario in which each sub-task completes in a single tool invocation. This abstraction allows us to highlight the logical sequence of steps without delving into low-level agentic behaviors.
%It is worth noting that the considered scenario refers to a simplified case, where each sub-task is assumed to be completed within a single invocation of the corresponding tool. 

\begin{algorithm}[h]
\textbf{Input:} Infrastructure file, Natural language description of the policies \\
\textbf{Output:} Verified Infrastructure file, Generated \texttt{Rego} rules.

\begin{algorithmic}[1]
\State Retrieve \emph{OPA} Knowledge using RAG Tool
\State Generate \texttt{Rego} Rules from the natural language description
\State Verify the \texttt{Rego} Rules using the Checker Tool
\If{Rules are wrong}
    \State \textbf{go to} 2
\EndIf
\State Preprocess the Infrastructure file using the Terraform Tool
\State Validate the Infrastructure JSON file against the \texttt{Rego} Rules using the Policy Validation Tool
\If{Infrastructure is not policy-compliant}
    \State Retrieve Terraform Knowledge using RAG Tool
    \State Correct the Infrastructure file
    \State \textbf{go to} 7
\EndIf
\State \Return Verified Infrastructure file, Generated \texttt{Rego} rules.
\end{algorithmic}
\caption{Expected \toolname{} workflow}
\label{alg:example}
\end{algorithm}

In real cases, the agentic nature architecture of \toolname{} often requires multiple iterations  with the same tool in order to iteratively refine the output and achieve  satisfactory results.
However, this  complexity is abstracted away from the end user.
\toolname{} manages all intermediate decisions and tool invocations internally. This design ensures a seamless experience, allowing users to focus on high-level objectives while the system autonomously orchestrates the underlying reasoning and execution processes.

%The agent autonomously manages all intermediate tool calls and adjustments, ensuring that the overall user experience remains seamless and consistent regardless of how many tool invocations are required behind the scenes. This abstraction enables users to focus on high-level goals, while \toolname{} takes care of orchestrating the underlying decision-making and execution process.

\subsection{Running Example}
%In response, the system automatically generates the appropriate \texttt{Rego} rule. Since the infrastructure satisfies the specified constraint, \toolname{} verifies compliance with the desired policy without requiring any modifications to the existing Terraform configuration.

\begin{figure}[bt]
    \centering
\begin{tcolorbox}[colback=NavyBlue!5!white,colframe=NavyBlue!70!black,title=\textbf{Terraform File excerpt},
    bottom=0pt,
    boxsep=4pt
]
\vskip-10pt
\begin{lstlisting}[basicstyle=\ttfamily\scriptsize]
...
resource "proxmox_virtual_environment_vm" "cloned_vm" {
  ...
  cpu {
    cores = 4
  }
  memory {
    dedicated = 8192
  }
  disk {
    interface    = "scsi0"         
    datastore_id = "Storage"   
    size         = 150
  }
  network_device {
    bridge = "intVM"
    model  = "virtio"
  }
  tags        = ["PaC"]
  description = "Cloned for PaC test"
}
\end{lstlisting}
\end{tcolorbox}
%\vskip-10pt
\begin{tcolorbox}[
    colback=Yellow!5!white,
    colframe=Yellow!70!black,
    title=\textbf{Policy prompt},
    bottom=0pt,
    boxsep=4pt
]
Allow only virtual machines with 4 cores in Terraform.
\end{tcolorbox}

\begin{tcolorbox}[
    colback=ForestGreen!5!white,
    colframe=ForestGreen!70!black,
    title=\textbf{Generated \texttt{Rego} rule},
    bottom=0pt,
    boxsep=4pt
]
\vskip-10pt
\begin{lstlisting}[basicstyle=\ttfamily\scriptsize]
package terraform

deny[msg] if {
    resource := input.planned_values.root_module.resources[_].values
    cpu := resource.cpu[_].cores
    cpu != 4
    msg := "VM must have exactly 4 cores"
}
\end{lstlisting}
\end{tcolorbox}
\caption{\toolname{} running example}
\label{fig:running_example}
\end{figure}

A minimal yet illustrative example of the capabilities of \toolname{} is shown in Figure~\ref{fig:running_example}. The environment consists of a single machine equipped with 4 CPU cores and 8~GB of RAM. The user instructs \toolname{} to allow only machines with exactly 4 cores.
\toolname{} processes this request by generating the corresponding \texttt{Rego} policy, then evaluating the compliance of the infrastructure against it. Since the current configuration satisfies the constraint, the system confirms policy compliance without requiring any modifications to the Terraform definition.

\section{Experimental Results}
\label{sec:results}
The experimental evaluation conducted aimed to assess the effectiveness of \toolname{} in the generation and validation of Policy as Code, with a focus on its applicability in a Terraform-based IaC scenario.
The evaluation considered the following key aspects:
\begin{itemize}
    \item \textbf{Syntactic correctness} of the generated \texttt{Rego} policies;
    \item \textbf{Semantic alignment} of the policies with the natural language user instructions;
    \item \textbf{Detection capability} for identifying policy violations within Terraform execution plans;
    \item \textbf{Repair effectiveness} in automatically correcting non-compliant infrastructure definitions.
\end{itemize}
All experiments were conducted on a ProxMox-based Asus ESC4000A-E12 server with an AMD EPYC 9004 processor, 2x 48GB L40s NVIDIA GPUs, and 196 GB of RAM. 
The publicly available LLMs were run using Ollama inside a Ubuntu 24.04 virtual machine hosted on the server, provided with 16 cores, 128 GB of RAM, and both available GPUs.
Closed-source models were accessed via API-based interactions using their publicly available endpoints. %The interaction with the closed models occurred using API calls.

\subsection{Evaluation methodology}
We evaluated \toolname{} on a defined and fixed Terraform infrastructure, with five distinct policy prompts of increasing difficulty. Among these, three  prompts required the modification of the provided infrastructure, due to their incompatibility with the original IaC definition.

%To demonstrate the effectiveness of our approach, we also performed an ablation study, repeating the same tests in a scenario leveraging only the LLM, without any external support, and in a scenario where the LLM is provided with the external RAG knowledge, but does not have access to the tools.
To assess the effectiveness of each system component, we conducted an ablation study with the following configurations:
\begin{itemize}
    \item \textbf{LLM-only:} the base LLM is used without access to retrieval or external tools;
    \item \textbf{LLM + RAG:} the model is enhanced with retrieval capabilities but lacks access to tool execution;
    \item \textbf{\toolname{} (full):} the complete agentic system with both RAG and tool invocation capabilities enabled.
\end{itemize}

Furthermore, we tested different LLMs for each of the discussed scenarios, to understand to what extent the capabilities of the ``raw'' LLM affect the overall generation performance of the system. We chose to evaluate \toolname{} with three different models: the open-weight model \texttt{Qwen3} in its 30 billion parameters version and the closed \texttt{GPT-4o} and \texttt{Claude Sonnet 4} models. These models represent the current state-of-the-art and support tool calling, ensuring a fair comparison in all the described scenarios.

\subsection{LLM vs RAG vs Agentic RAG}
In table~\ref{tab:performance} are depicted the results of the ablation study. The table shows a comparison across three different approaches, LLM, RAG, and Agentic RAG, applied to three models (\texttt{Qwen3:30b}, \texttt{GPT-4o}, and \texttt{Claude Sonnet 4}). The \textit{Model} column specifies the large language model employed for the specific batch of tests. The \textit{Configuration} column indicates whether the base LLM, RAG, or agentic RAG configuration was used. The \textit{Syntax} column shows how many out of five \texttt{Rego} policy generations were syntactically correct, and the \textit{Semantic} column reports how many of the syntactically correct policies were also semantically correct. The last column, \textit{Notes}, provides additional observations regarding each setup, such as errors encountered or behavior noted during the executions.
As expected, the agentic approach used in \toolname{} greatly enhances the system's capability of generating syntactically and semantically correct \texttt{Rego} policies. 
\vskip-10pt
\begin{table}[h]
\centering
\small
\renewcommand{\arraystretch}{1.3}
\begin{tabular*}{\columnwidth}{@{\extracolsep{\fill}}ll>{\centering}p{1.5cm}>{\centering}p{1.8cm}p{4cm}@{}}
\toprule
\textbf{Model} & \textbf{Configuration} & \textbf{Syntax} & \textbf{Semantic} & \textbf{Notes} \\
\midrule
\texttt{Qwen3:30b} & LLM & 0/5 & --- & \raggedright\arraybackslash \texttt{Rego} rules generated with syntactical errors \\
\addlinespace[0.2em]
\texttt{Qwen3:30b} & RAG & 0/5 & --- & \raggedright\arraybackslash \texttt{Rego} rules generated with syntactical errors \\
\addlinespace[0.2em]
\texttt{Qwen3:30b} & Agentic RAG & 4/5 & 4/5 & \raggedright\arraybackslash Loop during the policy correction for 1 prompt \\
\midrule
\texttt{GPT-4o} & LLM & 0/5 & --- & \raggedright\arraybackslash \texttt{Rego} rules generated with syntactical errors \\
\addlinespace[0.2em]
\texttt{GPT-4o} & RAG & 0/5 & --- & \raggedright\arraybackslash \texttt{Rego} rules generated with syntactical errors \\
\addlinespace[0.2em]
\texttt{GPT-4o} & Agentic RAG & 5/5 & 5/5 & \raggedright\arraybackslash 1/3 Terraform file modified \\
\midrule
\texttt{Claude Sonnet 4} & LLM & 5/5 & 5/5 & \raggedright\arraybackslash \texttt{Rego} rules generated without external knowledge \\
\addlinespace[0.2em]
\texttt{Claude Sonnet 4} & RAG & 5/5 & 5/5 & \raggedright\arraybackslash \texttt{Rego} rules generated \\
\addlinespace[0.2em]
\texttt{Claude Sonnet 4} & Agentic RAG & 5/5 & 5/5 & \raggedright\arraybackslash \texttt{Rego} rules generated and checked \\
\bottomrule
\end{tabular*}
\vskip3pt
\caption{Performance summary and comparison}
\label{tab:performance}
\end{table}
\vskip-25pt
Base LLMs most likely lack knowledge about OPA and \texttt{Rego} to effectively generate correct policies. Furthermore, even when such knowledge is provided through the RAG module, most of the models are not capable of generating satisfying policies on the first try, failing the syntax check.
Hence, the availability of deterministic verification tools is crucial to allow the system to iteratively correct itself, eventually reaching a satisfying output.
Our approach achieves success most of the time, with failures related to an excessive number of retries during the policy correction (in the tests, the max amount of workflow iterations was fixed to 3) or the system's inability to modify the original Terraform file.

Surprisingly, the \texttt{Claude Sonnet 4} model already possesses the required knowledge and is capable of generating correct \texttt{Rego} rules for our tests even in the base LLM scenario, showing how powerful this cutting-edge model is. However, it is worth noting that only the agentic approach ensures that the generated rules are syntactically and semantically correct, while the simple use of an LLM does not provide any warranty in that sense. This is a consequence of the agentic approach's capacity to reason about any errors and implement appropriate corrections autonomously.

Furthermore, another crucial result obtained involves the use of smaller and cheaper models. Despite a very powerful (and costly) model, such as \texttt{Claude Sonnet 4} might be able to generate satisfying rules, \toolname{}'s approach enables the use of much smaller models (30 billion parameters in our tests) to achieve comparable results.
This allows the use of one of the appropriate publicly available models in the scenario of PaC generation and IaC compliance verification, effectively eliminating dependencies on external model providers.

%\vskip -10pt
\subsection{Model Comparison for the Agentic RAG}
%Considering only the Agentic RAG configuration, we observed that, as expected, the capabilities of the underlying LLM have an impact on the generation performances. However, such impact involves mainly the required number of RAG and tool calls required to accomplish the task. 
In the full Agentic RAG configuration of \toolname{}, we evaluated how the choice of underlying LLM affects overall performance. As expected, the model's capabilities significantly influence task execution, primarily in terms of the number of RAG and tool invocations required to reach a successful outcome.
As reported in Table~\ref{tab:rag-tool-calls},  more powerful models tend to require fewer (average) calls to the tools to produce satisfying policies and to correct the infrastructure file.
However, this reduction comes at a cost, as more powerful models are also more expensive to use, while the involved tools have negligible costs. % \revfr{(tranne per il manual check delle regole)}.
Thus, the choice of LLM in \toolname{} should balance between the capabilities and the cost of the executions.
\vskip-10pt
\begin{table}[ht]
\centering
\small
\renewcommand{\arraystretch}{1.3}
\begin{tabular}{l>{\centering}p{3cm}>{\centering\arraybackslash}p{3cm}}
\toprule
\textbf{Model} & \textbf{RAG Call (avg)} & \textbf{Tool Call (avg)} \\
\midrule
\texttt{Qwen3:30b} & 4.4 & 11.4 \\
\texttt{GPT-4o} & 3.8 & 9.0 \\
\texttt{Claude Sonnet 4} & 3.2 & 7.8 \\
\bottomrule
\end{tabular}
\vskip3pt
\caption{Average RAG and tool calls in the Agentic RAG}
\label{tab:rag-tool-calls}
\end{table}
\vspace*{-3.5em}
\section{Conclusions and Future Work}
\label{sec:conclusions}

In this work, we introduced \toolname{}, a novel agentic system for the generation of Policy as Code  and the validation of Infrastructure as Code  configurations.  
By adopting the Agentic RAG paradigm, \toolname{} provides an effective solution for the effortless creation and validation of security policies in IaC environments.
Our results demonstrate that \toolname{}, thanks to the use of deterministic validation tools, significantly improves performance, enabling accurate and reliable policy implementation, particularly when using smaller language models.
Among the models evaluated, \texttt{Claude Sonnet 4} emerged as the most effective for this task. However, the adoption of the agentic RAG paradigm also enables the effective use of smaller models (e.g. \texttt{Qwen3:30b}), which, when supported by a well-curated knowledge base and appropriate tools, can achieve performance comparable to that of larger and more expensive models.
In future work, we plan to extend \toolname{} with automated semantic verification of the generated \texttt{Rego} rules. This remains a challenging task due to the complexity of verifying semantic correctness automatically. Additionally, we aim to explore the integration with self-RAG~\cite{asai2023selfraglearningretrievegenerate} technologies, which could further enhance the autonomy and adaptability of the system during the generation and validation phases.

\begin{credits}
\subsubsection{\ackname} 
This work was partially supported by the SERICS project \hfill
\\(PE00000014) under the MUR National Recovery and Resilience Plan funded by the European Union - NextGenerationEU.
The work of Francesco A. Pironti was supported by Agenzia per la cybersicurezza nazionale under the 2024-2025 funding program for promotion of XL cycle PhD research in cybersecurity (CUP H23C24000640005).

%\subsubsection{\discintname}
%It is now necessary to declare any competing interests or to specifically state that the authors have no competing interests. Please place the statement with a bold run-in heading in small font size beneath the (optional) acknowledgments, for example: The authors have no competing interests to declare that are relevant to the content of this article. Or: Author A has received research grants from Company W. Author B has received a speaker honorarium from Company X and owns stock in Company Y. Author C is a member of committee Z.
\end{credits}
\bibliographystyle{unsrt}
\bibliography{refs}
\end{document}